%% file: paper_la.tex
\documentclass[twocolumn,showpacs,aps,prl,superscriptaddress]{revtex4}

\usepackage{graphicx}
\usepackage{dcolumn}
\usepackage{amsmath}
\usepackage{epsfig}

\input pubboard/babarsym

\newcommand{\BABARPubYear}    {02}
\newcommand{\BABARPubNumber}  {06}

\newcommand{\SLACPubNumber} {9261}

\def\figurebox#1#2#3{
    \def\arg{#3}
    \ifx\arg\empty
    {\hfill\vbox{\hsize#2\hrule\hbox to #2{\vrule\hfill\vbox to #1{\hsize#2\vfill}\vrule}\hrule}\hfill}
    \else
    {\hfill\epsfbox{#3}\hfill}
    \fi}

\begin{document}

\preprint{\babar-PUB-\BABARPubYear/\BABARPubNumber} 
\preprint{SLAC-PUB-\SLACPubNumber} 

\begin{flushleft}
\babar-PUB-\BABARPubYear/\BABARPubNumber\\
SLAC-PUB-\SLACPubNumber\\
\end{flushleft}

\title{
{\large \bf A Measurement of the $\Bz \to \jpsi\pi^+\pi^-$ Branching Fraction} 
}

\input pubboard/authors_0206

\date{\today}

\begin{abstract}
We present a measurement of the branching fraction for the decay of
the neutral \B\ meson into the final state $\jpsi\pipi$.  The data set
contains approximately 56 million \BB\ pairs produced at the $\FourS$
resonance and recorded with the \babar\ detector at the \pep2\ 
asymmetric-energy \epem\ storage ring. 
The result of this analysis is $\mathcal{B}$($\Bz\to\jpsi\pipi$) = 
(4.6 $\pm$ 0.7 $\pm$ 0.6)$\times 10^{-5}$, where the first error is 
statistical and the second is systematic.  In addition we measure 
$\mathcal{B}(\Bz\to\jpsi\rho^0) = (1.6 \pm 0.6 \, \pm 0.4\,)\times 10^{-5}$.
\end{abstract}


\maketitle
\label{sec:Introduction}

In the Standard Model, the decay $\Bz\to\jpsi\rho^0$ can give rise to 
\CP-violating asymmetries (directly and through $\Bz$-$\Bzb$ mixing). 
Therefore it is interesting to study the decay mode $\Bz\to\jpsi\pipi$ to 
understand the $\jpsi\rho^0$ component in the final state. Since these decays 
are Cabibbo and color suppressed, they are a sensitive probe of rare and 
exotic physics processes, such as penguin contributions and box diagrams 
containing charged Higgs bosons. These effects may appear as deviations of the
branching fraction from the Standard Model prediction of 
$\mathcal{B}(\Bz\to\jpsi\pipi) = (4.6\pm 0.8)\times 
10^{-5}$~\cite{Prediction}.  This decay mode has not previously been observed. 
CLEO quotes an upper limit of $\mathcal{B}(\Bz\to\jpsi\rho^0) < 
2.5\times 10^{-4}$ at the 90\% confidence level~\cite{Bishai:1995yj}.  Here 
we present the first measurement of $\mathcal{B}(\Bz\to\jpsi\pipi)$.

\label{sec:babar}
The data used in the present analysis were collected
at the \pep2\ storage ring with the \babar\ detector, described in detail
elsewhere~\cite{ref:babar}.  
Charged particles are detected, and their momenta measured, with a 
40-layer drift chamber (DCH) and a five-layer silicon vertex 
tracker (SVT), both operating in a 1.5\,T solenoidal magnetic field.  
Surrounding the DCH is a detector of internally reflected Cherenkov 
radiation (DIRC), and outside this is a CsI(Tl) electromagnetic calorimeter 
(EMC).  The iron flux return of the solenoid is instrumented with resistive
plate chambers (IFR).  
The data sample used for the analysis contains approximately 56 million \BB\
pairs, corresponding to a luminosity of 51.7\invfb\ recorded near the \FourS\ 
resonance.  An additional 6.4\invfb, recorded approximately 40\mev\ below the 
\FourS\ peak, were used to study continuum backgrounds.

Events containing \BB\ pairs are selected based on track multiplicity and 
event topology~\cite{Aubert:2001xs}.  At least three tracks are required to
originate near the nominal beam spot, with polar angle in the range
$0.41 < \theta_{\rm lab} < 2.54$\,\rad, transverse momentum 
greater than 100\mevc, and a minimum number of DCH hits used in the track fit. 
To reduce continuum
background the ratio of second to zeroth Fox-Wolfram moment, $R_2 = H_2/H_0$, 
is required to be less than 0.5.  The sum of charged and neutral energy must 
be greater than 4.5\gev\ in the laboratory frame.  The primary vertex of the 
event must be within 0.5\cm\ of the average measured position of the 
interaction point in the plane transverse to the beamline.  

The $\jpsi$ is reconstructed in the $\epem$ and $\mumu$ final states.  
Electron candidates must satisfy the requirement that the ratio of 
calorimeter energy to track momentum lies in the range $0.75 < E/p < 1.3$, 
the cluster shape and size are consistent with an electromagnetic shower, and 
the energy loss in the DCH is consistent with that for an electron.  
If an EMC cluster close to the electron track is consistent with originating 
from a bremsstrahlung photon, it is combined with the electron candidate.   

Muon candidates must satisfy requirements on the number of interaction lengths
of IFR iron penetrated ($N_{\lambda}>2$), the difference between the measured
and expected interaction lengths penetrated 
($|N_{\lambda}-N^{\rm exp}_{\lambda}|<2$), the position match between the 
extrapolated DCH track and the IFR hits, and the average and spread of the 
number of IFR strips hit per layer.

Pion candidates are accepted if they originate from close to the beam spot and
are not consistent with being a kaon.  The algorithm uses \dedx\ 
information from the SVT and DCH, and the Cherenkov angle and number of 
photons from the DIRC.  

Tracks are required to lie in polar-angle ranges where particle
identification efficiency is measured with known control samples.
The allowed ranges are
$0.41 < \theta_{\rm lab} < 2.41$\,\rad for electrons, 
$0.30 < \theta_{\rm lab} < 2.70$\,\rad for muons, and 
$0.35 < \theta_{\rm lab} < 2.50$\,\rad for pions, which correspond 
approximately to the geometrical acceptances of the EMC, IFR, and DIRC, 
respectively.

Identified electron and muon pairs are fit to a common vertex and
must lie in the $\jpsi$ invariant mass interval 2.95 (3.06) to 3.14\gevcc for 
the $\epem$ ($\mumu$) channel.

\begin{figure}[htb]
\begin{center}
\includegraphics[width=0.9\linewidth]{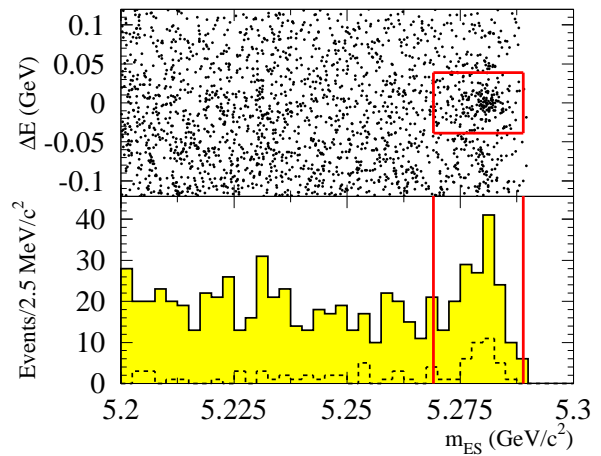}
\caption{Signal for $\Bz\to\jpsi\pipi$.  The upper plot shows the distribution
of events in the $\Delta E$-$m_{\rm ES}$ plane, where the box represents the 
final selection criteria.  The lower plot shows the distribution in 
$m_{\rm ES}$ of events with $| \Delta E | < 39\mev$, where the dashed (solid)
line corresponds to events in the $\KS$ (non-$\KS$) region in $M(\pipi)$ 
($0.45$-$0.55\gevcc$).  The vertical lines represent the final selection.}
\label{fig:mesde}
\end{center}
\end{figure}

\Bz\ candidates are formed by combining a $\jpsi$ candidate with a pair of 
oppositely-charged pion candidates consistent with coming from a common 
decay point.  We also require the positions of the vertices of the lepton pair
and the pion pair to be consistent.  Further selection requirements 
are made using two kinematic variables: the difference, $\Delta E$,
between the energy of the candidate and the beam energy $E^{\rm cm}_{\rm beam}$
in the center-of-mass frame, and the beam-energy substituted mass, 
$m_{\rm ES} = \sqrt{(E^{\rm cm}_{\rm beam})^2 - (p^{\rm cm}_B)^2}$.  After
applying the loose requirements $5.2 < m_{\rm ES} < 5.3\gevcc$ and 
$| \Delta E | < 0.12\gev$, approximately one-quarter of events contain more 
than one \Bz\ candidate, from which we keep the one with the smallest 
$| \Delta E |$.  The distribution of the candidates in $\Delta E$ and 
$m_{\rm ES}$ is shown in Fig.~\ref{fig:mesde}.  For the final signal sample, 
we require $| m_{\rm ES} - 5279.0\mevcc | < 9.9\mevcc$ and $| \Delta E | < 
39\mev$, which correspond to $4\sigma$ and $3\sigma$ ranges
in the resolutions for $m_{\rm ES}$ and $\Delta E$.  After all selection 
criteria have been applied, 213 events remain.

An unbinned, extended maximum-likelihood~\cite{Barlow:1990vc} fit is performed
on the invariant mass distribution of the two pions for the selected events, to
determine the various contributions to the $\Bz\to\jpsi\pipi$ events.
We consider five categories: (i) $\Bz\to\jpsi\rho^0$ events; (ii) 
$\Bz\to\jpsi\KS (\KS\to\pipi)$ events; (iii) $\Bz\to\jpsi\pipi$ (non-$\rho^0$ 
signal) events; (iv) background from events without a real $\jpsi$; (v) 
inclusive-$\jpsi$ background from events containing a real $\jpsi$.
A probability density function (PDF) is constructed for each of these five 
cases.  The total PDF is then formed from the sum of the five PDFs and fit 
to the data.  The $\Bz\to\jpsi\KS$ mode is not considered to be signal 
for the purposes of determining the branching fraction for $\Bz\to\jpsi\pipi$.
  
\label{sec:pdfs1}
The PDF used to model the $\Bz\to\jpsi\rho^0$ mode is a relativistic $P$-wave
Breit-Wigner function~\cite{PisutRoos}:\\
$F_{\rho}(m) = (m \, \Gamma(m) \, P^{2L_{\rm eff}+1})/((m_{\rho}^2-m^2)^2+
m_{\rho}^2\Gamma(m)^2),\\
{\rm where} \hspace{0.3cm} \Gamma(m) = \Gamma_0 (\frac{q}{q_0})^3 
(\frac{m_{\rho}}{m}) (\frac{1 + R^2 q_0^2}{1 + R^2 q^2}).$ 
$q(m)$ is the pion momentum in the di-pion rest frame, with $q_0 = 
q(m_{\rho})$.  $m\equiv M(\pipi)$ is the two-pion invariant mass and $P$ is 
the $\jpsi$ momentum in the $\Bz$ rest frame.  $m_{\rho} = 770\mevcc$, 
$\Gamma_0 = 150\mevcc$, and $m_{\pi} = 140\mevcc$.  $L_{\rm eff}$ is the 
effective orbital angular momentum between the $\jpsi$ and the $\rho^0$, which 
can take any value between 0 and 2 and so is allowed to float in the fit.  $R$ 
is the Blatt-Weisskopf barrier-factor radius~\cite{Blatt}.  The fit is 
performed with $R$ equal to two values (0.5 and 1.0\,fm) and the results of 
the two fits are averaged.  

The PDF for the $\Bz\to\jpsi\KS$ mode is a single Gaussian function with the 
mass and width fixed to values obtained by fitting a sample of simulated 
$\jpsi\KS$ events.  Allowing these parameters to vary in the final $M(\pipi)$ 
fit does not change the results.

The PDF used to model the $\Bz\to\jpsi\pipi$ (non-$\rho^0$ signal) contains a
three-body phase space factor $q(m)P(m)$ and a factor of $P(m)^2$
motivated by angular momentum conservation: $F_{\rm ph}(m) = q(m)P(m)^3$. If 
the $\pipi$ is in an $S$-wave, angular momentum conservation results in a
factor of $P(m)^2$, while a $D$-wave yields the second power of $P(m)$ or
higher.

\label{sec:pdfs2}
The PDF for the $M(\pipi)$ distribution for background events without a real 
$\jpsi$ is derived from a fake-$\jpsi$ sample selected in data as described
above except that at least one of the lepton candidates 
must fail the appropriate particle identification requirements.  A Monte Carlo
study confirms that the $M(\pipi)$ distribution obtained with this procedure 
correctly describes the shape of the non-$\jpsi$ background.  The resulting 
distribution is parametrized using the sum of two Weibull 
functions~\cite{Weibull} and a Breit-Wigner.  The Breit-Wigner describes the 
$\rho^0$ component of the non-$\jpsi$ background.

The PDF for the $M(\pipi)$ shape for background events containing a real 
$\jpsi$ is obtained from a simulated $\B\to\jpsi\X$ sample
equivalent to a luminosity of 81\invfb.  Events in which the system $X$ is 
\pipi (non-resonant), $\rho^0$, or \KS(\pipi)\ are removed from the sample.  
The resulting shape is described by a Weibull function.

The normalization of the background components is obtained from
samples in data and simulation.  The level of non-$\jpsi$ background is 
obtained from sidebands of the $\jpsi$ mass distribution in data.  The 
$m_{ES}$ distribution for these sideband candidates is then fit to an ARGUS 
function~\cite{Albrecht:1990cs} to determine how many events pass the final 
selection criterion.  Scaling to the equivalent background in the 
$\jpsi$ mass region, using an exponential to describe the background shape
in the $\jpsi$ mass distribution, the expected non-$\jpsi$ background is found
to be 35.7$\pm$1.2 events.

The level of inclusive-$\jpsi$ background is obtained from the 
distribution of $m_{ES}$ for events in the $\Delta E$ signal region in both 
data and simulation.  In each case the $m_{ES}$ distribution is parametrized 
by a Gaussian function (to represent signal or peaking background) and an 
ARGUS function.  Peaking background originates from $\B\to\jpsi\X$ decays such
as $\B\to\jpsi\Kstar$, $\Bp\to\jpsi\rho^+$, and $\B\to\jpsi K_1$, that 
accumulate near $m_{ES}=5.279\,\gevcc$.

The non-peaking component of the inclusive-$\jpsi$ background is determined by 
subtracting the non-$\jpsi$ contribution, on the basis of the scaled sideband 
events described above, from the total ARGUS background in data.  The peaking 
component is determined from the Gaussian part of the $m_{ES}$ distribution in
$\B\to\jpsi\X$ simulation, where events with $X=\pipi$ (non-resonant), $\rho^0$
and $\KS (\pi^+\pi^-)$ have been removed.  The sum of peaking and non-peaking
components of the inclusive-$\jpsi$ background is found to be 61$\pm$11 events,
of which the peaking component comprises 6 events.  
Thus any associated uncertainties, such as branching fractions 
used in the $\jpsi X$ simulation, will not contribute significantly to the 
final systematic uncertainty.

The branching fraction is obtained from
\begin{equation}
\label{eqn:bf}
\mathcal{B}(\Bz\to\jpsi\pi^+\pi^-) = 
\frac{N_{\jpsi\pi\pi}}{N_{\Bz}\times\epsilon_{\jpsi\pi\pi}\times\mathcal{B}
(\jpsi\to\ellell)},
\end{equation}
where $N_{\jpsi\pi\pi}$ is the total signal yield obtained from the fit, 
$N_{\Bz}$ is the total number of \Bz and \Bzb in the 
data sample~\cite{Aubert:2001xs}, and $\epsilon_{\jpsi\pi\pi}$ is the signal 
efficiency.  The $\jpsi$ branching fraction $\mathcal{B}(\jpsi\to\ellell)$ is 
fixed to 11.81\%~\cite{Groom:in}.  We assume that the branching fraction for 
$\FourS\to\Bz\Bzb$ is one-half.

The signal efficiencies for all requirements apart from particle  
identification criteria are derived from simulation.  Lepton and pion 
identification efficiencies are determined with samples of known muons, 
electrons and pions in the data from the following processes: $\mumu\g$, 
$\mumu\epem$, \epem, $\epem\g$, $\Dstarp\to\Dz\pip$ ($\Dz\to\Km\pip$), and 
\KS\to\pip\pim.  The efficiencies are determined as a function of momentum, 
and polar and azimuthal angle.  We find $\epsilon(\jpsi\rho^0)=(27.1\pm0.3)\%$
and $\epsilon(\jpsi\pipi,$ non-resonant$)=(27.0\pm0.3)\%$.
The final corrected signal efficiency of ($27.1 \pm 0.2$)\% is taken as the 
average of the $\jpsi\rho^0$ and $\jpsi\pipi$ (non-resonant) efficiencies,
where the error is from Monte Carlo statistics.  

\begin{figure}[htb]
\begin{center}
\includegraphics[width=0.9\linewidth]{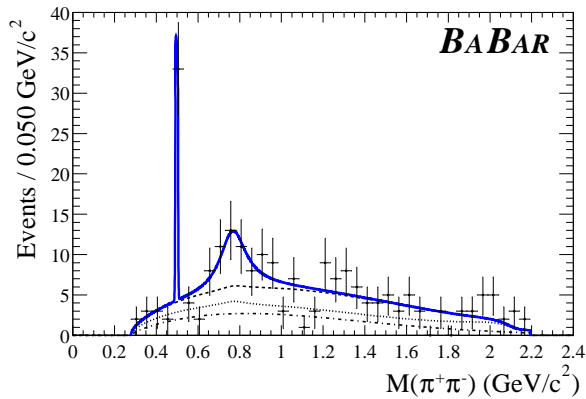}
\caption{Distribution of the invariant mass $M(\pipi)$ for events passing all 
selection criteria. The solid line is the result of the unbinned likelihood 
fit. The dashed line represents the sum of background and non-$\rho^0$ signal 
components.  The dotted (dot-dashed) line shows the total (inclusive-$\jpsi$) 
background.  The spike corresponds to $\Bz\to\jpsi\KS$ events.}
\label{fig:fit}
\end{center}
\end{figure}

\label{sec:Physics}
A likelihood fit is performed on the $M(\pipi)$ distribution in data with the 
normalization of the non-$\jpsi$ background fixed to 35.7 events and 
the inclusive-$\jpsi$ background to 61.  Thus only the yields for 
$\jpsi\rho^0$, $\jpsi\pipi$ (non-$\rho^0$ signal),  and $\jpsi\KS$ events are 
allowed to vary.  The results of the fit are overlaid on the data points in
Fig.~\ref{fig:fit}.  The goodness-of-fit $\chi^2$ is 33.4 for 38 data points.

The result of the fit is $84\pm13$ signal events, of which $28\pm10$ 
are in the $\rho^0$ component and $55\pm15$ are in the non-$\rho^0$ signal 
component.  The number of events in the $\KS$ component is $28\pm5$.  
Inserting the result into Eq.~\ref{eqn:bf} yields the branching fraction
$\mathcal{B}(\Bz\to\jpsi\pipi) = (4.6 \pm 0.7)\times 10^{-5}$,
where the error is statistical.  

The signal yield can be checked by counting the number of
events passing all the selection criteria and subtracting the estimated 
numbers of background and $\jpsi\KS$ events.  This method gives 
$87 \pm 15$ $\jpsi\pi^+\pi^-$ events, where the error is statistical.

\label{sec:Systematics}
The systematic errors on the final branching fraction measurement arise from
uncertainties on the signal efficiency, fitted yield, number of \BB\ pairs
produced, and $\jpsi\to\ellell$ branching fraction.  $N_{\BB}$ is known to 
$1.1\%$ with the dominant contribution to the uncertainty coming from the error
on the efficiency for the $\Bz\Bzb$ selection.  
$\mathcal{B}(\jpsi\to\ellell)$ is known to 1.7\% (fractional)~\cite{Groom:in}.

The uncertainty on the pion identification efficiency is 1.8\% per pion. 
Contributions to this error come from the limited size of the data sample used
to determine the efficiency, the uncertainty on the kaon contamination in the 
sample, and residual differences between the efficiencies for the known pions 
in data and for pions from $B^0\to\jpsi\pipi$, determined from Monte Carlo 
simulation.

Uncertainties on electron and muon particle identification efficiencies come 
from studies using $B\to\jpsi X$ events in data.  Fits to the $M(\jpsi)$ 
distribution in these events, under different selection criteria, give 
estimates of the electron and muon identification efficiencies and their 
errors.  The overall error for the identification criteria used in this 
analysis is 1.3\%.

The uncertainty on the determination of the tracking efficiency is 1.3\% per 
track, and is summed for the four tracks from the \Bz decay.  The 
efficiency of the convergence requirement on the $\pipi$ vertex fit has been 
studied with a sample of $\psitwos\to\ellell$ decays. Data and simulation are 
found to be in good agreement, with an associated systematic error of 1\%.  
The unknown $\rho^0$ helicity in the $\jpsi\rho^0$ component of the final 
sample introduces a systematic error on the efficiency of 2.5\%, as determined
from the efficiency variation between simulated samples with helicity 0 and 1.
The limited amount of simulated data leads to an uncertainty in signal 
efficiency of 0.7\%.
To determine the effect of the signal and background shapes and the 
background yields on the fitted yields, the fixed parameters of these PDFs are
varied within their uncertainties, allowing for correlations.  This produces a
total systematic error due to fit parameter variation of 9.7\%, which is 
dominated by the errors in the background yields.  
The final fit neglects resonances such as $f_0(980), f_2(1270)$ 
and $\rho^0(1450)$.  Allowing for the addition of such terms in the 
likelihood function results in a systematic uncertainty on the yield of 2.1\%. 
The total systematic uncertainty from all sources is found to be 12.3\%.

\label{checks}
The analysis is repeated with variations in the selection criteria.  Taking 
into account statistical correlations between the results, we find that 
variations are consistent with statistical fluctuations due to
the addition or removal of some of the events in the sample.

The branching fraction can be measured separately for events containing a 
$\jpsi\to\epem$ or a $\jpsi\to\mumu$ candidate.
The results from these subsamples are $\mathcal{B}(\Bz\to\jpsi\pipi)_{ee} = 
(5.3 \pm 1.1)\times 10^{-5}$ and $\mathcal{B}(\Bz\to\jpsi\pipi)_{\mu\mu} = 
(4.0 \pm 1.0)\times 10^{-5}$, where the errors are statistical.

Another way to model the backgrounds is to use a smoothing algorithm on the
simulated $\B\to\jpsi\X$ and fake-$\jpsi$ data events, rather than 
impose a parametrization.  The resulting PDFs follow fluctuations and check 
how strongly the fitted signal yields depend on the chosen method of 
describing the backgrounds.  Changing the background modeling
in this way alters the total fitted yield by less than one event.

The $M(\pipi)$ distribution shows a clear peak at the $\rho^0$ mass.
The fit result of $28\pm10$ events for the $\rho^0$ signal leads to a 
branching fraction of $\mathcal{B}(\Bz\to\jpsi\rho^0) = (1.6 \pm 0.6 \,
({\rm stat}) \pm 0.4 \,({\rm syst}))\times 10^{-5}$.  The systematic error
includes a contribution from the effect of using an alternative PDF to 
describe the non-$\rho^0$ signal.  The shape is from a polynomial fit to data 
recorded in $\pi$-$\pi$ scattering experiments~\cite{Gaspero:gu} and thus 
provides an empirically-derived shape, in contrast to the default non-$\rho^0$ 
signal PDF, which is based on a phase-space assumption.  The assumption that
the non-$\rho^0$ signal is predominantly $S$-wave, and therefore 
interference with the $\rho^0$ can be neglected, has been checked on data.  A 
significant $S$-wave contribution means that the leptons from the $\jpsi$ have 
a helicity angle distribution $\propto\sin^2(\theta_{\jpsi})$.  For events in 
data with $M(\pipi)>1.1\gevcc$, we subtract the helicity cosine distribution 
for events with $m_{ES}<5.27\mevcc$ from the distribution for events in the 
signal $m_{ES}$ region and find that the shape of the resulting distribution 
is consistent with $\sin^2(\theta_{\jpsi})$.  Interference between $S$- and 
$P$-wave signal components integrates out in the $M(\pipi)$ projection, as 
long as the acceptance is symmetric in the cosine of the di-pion helicity 
angle, $\theta_{\pi\pi}$.  Studies using simulated non-resonant $S$-wave 
events show that there is no significant odd component to the acceptance 
function in $\cos(\theta_{\pi\pi})$.  Consequently, there is no such 
interference contribution to the $\pipi$ mass distribution

In summary, the branching fraction for $\Bz$ meson decay to the final state 
$\jpsi\pipi$ has been measured for the first time. The result, 
$\mathcal{B}(\Bz\to\jpsi\pipi) = (4.6 \pm 0.7 \,({\rm stat}) \pm 
0.6 \,({\rm syst}))\times 10^{-5}$, is consistent with the Standard Model 
prediction~\cite{Prediction}.  In addition, the technique of fitting the 
$M(\pipi)$ distribution allows a measurement of the branching fraction for the
$\jpsi\rho^0$ component.  The result is  $\mathcal{B}(\Bz\to\jpsi\rho^0) =
(1.6 \pm 0.6 \,({\rm stat}) \pm 0.4 \,({\rm syst}))\times 10^{-5}$. 

\input pubboard/acknow_PRL.tex

\end{document}

%% file: pubboard/authors_0206.tex
%
\author{B.~Aubert}
\author{D.~Boutigny}
\author{J.-M.~Gaillard}
\author{A.~Hicheur}
\author{Y.~Karyotakis}
\author{J.~P.~Lees}
\author{P.~Robbe}
\author{V.~Tisserand}
\author{A.~Zghiche}
\affiliation{Laboratoire de Physique des Particules, F-74941 Annecy-le-Vieux, France }
\author{A.~Palano}
\author{A.~Pompili}
\affiliation{Universit\`a di Bari, Dipartimento di Fisica and INFN, I-70126 Bari, Italy }
% ...
\collaboration{The \babar\ Collaboration}
\noaffiliation

%% file: pubboard/acknow_PRL.tex
We are grateful for the excellent luminosity and machine conditions
provided by our \pep2\ colleagues, 
and for the substantial dedicated effort from
the computing organizations that support \babar.
The collaborating institutions wish to thank 
SLAC for its support and kind hospitality. 
This work is supported by
DOE
and NSF (USA),
NSERC (Canada),
IHEP (China),
CEA and
CNRS-IN2P3
(France),
BMBF and DFG
(Germany),
INFN (Italy),
NFR (Norway),
MIST (Russia), and
PPARC (United Kingdom). 
Individuals have received support from the 
A.~P.~Sloan Foundation, 
Research Corporation,
and Alexander von Humboldt Foundation.

%% file: paper_la.bbl
\begin{thebibliography}{99}
\bibitem{Prediction}
The world average values~\cite{Groom:in} for the 
branching fractions $\Bz\to\jpsi K^{*0}$ and $\Bz\to\jpsi K^+\pi^-$ have been
scaled by $0.5|V_{cd}|^2/|V_{cs}|^2$ and $|V_{cd}|^2/|V_{cs}|^2$, 
respectively.  The number quoted in the text is the sum of the two scaled 
values, and assumes that the tree amplitude is dominant.
\bibitem{Weibull}
The Weibull 
function can be written as\\
$W(m) = C\,V\,(m-M_{\rm on})^{(C-1)} \times \exp 
[-V (m-M_{\rm max})^C],\\
{\rm where} \hspace{0.3cm} V = (C-1)/(C(M_{\rm max}-M_{\rm on})^C)$.\\
$M_{\rm max}$ is the position of the function maximum, $M_{\rm on}$ is the 
lower kinematic cut-off, and $C$ is a general shape parameter.
\end{thebibliography}
